# Sensitivity of Remote Focusing Microscopes to Magnification Mismatch

S. Mohanan[1], A.D. Corbett[1]

1. Department of Physics and Astronomy, University of Exeter, EX4 4QL, UK

Correspondence to: A.D. Corbett. Email: a.corbett@exeter.ac.uk

## Summary

Remote focusing (RF) is a technique that greatly extends the aberration-free axial scan range of an optical microscope. To maximise the diffraction limited depth range in an RF system, the magnification of the relay lenses should be such that the pupil planes of the objectives are accurately mapped on to each other. In this paper we study the tolerance of the RF system to magnification mismatch and quantify the amount of residual spherical aberration present at different focusing depths. We observe that small deviations from ideal magnification results in increased amounts of residual spherical aberration terms leading to a reduction in the dynamic range. For high numerical aperture objectives, the simulation predicts a 50% decrease in the dynamic range for 1% magnification mismatch. The simulation has been verified against an experimental RF system with ideal and non-ideal magnifications. Experimentally confirmed predictions also provide a valuable empirical method of determining when a system is close to the ideal phase matching condition, based on the sign of the spherical aberration on either side of focus.

## Introduction

Live biological imaging requires acquisition of image volumes at high speed and high spatial resolution. One added constraint is that the rapid movement of the objective lens or the sample stage to refocus the microscope at different depths introduces vibrational artifacts. This can hinder the observation of transient biological phenomena. In addition to this, translating these relatively heavy components can reduce the temporal resolution of the system. It is then advantageous to decouple refocusing from the object space to a remote location in the optical train of the microscope.

Remote refocusing strategies include the introduction of passive optical elements into the optical path of the microscope. Multifocus Microscopes[1] use distorted diffraction gratings for simultaneous imaging of multiple planes in a single camera frame acquisition. These gratings are designed to compensate for spherical aberration introduced at specific depths and extends the axial ($z$) range of imaging using high numerical aperture (NA) objectives to a few tens of microns. Another refocusing method introduces a lenslet array into the optical path to form a Light Field Microscope[2]. Similar to diffraction gratings, this method allows for the capture of an entire volume in a single frame which eliminates refocusing time. However, they come with reduced flexibility in choosing the planes of interest within the sample volume as the optical elements are selected for a specific field of view. In addition to this, Light Field Microscopes have an inherent trade-off between the extended axial range of imaging and the spatial resolution of the microscope.

Other passive techniques include the use of phase masks to engineer the pupil function of the objective to reduce its sensitivity to defocus[3]. This method has been implemented along with a Light Sheet Fluorescence Microscope (LSFM) to scan through samples at 70 volumes per second (vps) with a ten-fold increase in the depth of field. As introducing a phase mask modifies the Optical Transfer Function of the objective, the images require post-processing using deconvolution techniques to retrieve the original spatial resolution. Active refocusing methods include variable focal length lenses such as Electro Tunable Lenses (ETL). ETLs have also been used along with LSFM for scanning across the focal volume at 30 vps[4]. The focal length of ETLs is adjusted to change the effective focal length of the objective to rapidly refocus the microscope. However, ETLs cannot compensate for high-NA defocus for a large range of $z$[5,6].

The "remote focusing" (RF) system proposed by Botcherby[7] allows for refocussing high-NA objectives at temporal resolutions only limited by camera speed. It can be easily combined with sectioning techniques such as confocal[8], light sheet[9], structured illumination[10] and multi-photon[11] microscopes to produce high contrast volumetric data. It also allows the selection of oblique planes to study features of interest within the sample volume[12,13] leading to increased flexibility in volumetric scanning. However, despite these advantages, the adoption of RF systems as a standard high-NA refocusing methodology has been slow due to high sensitivity to optical alignment of high NA objectives. There has been work done to help microscopists choose the best combination of lenses and the best alignment practices[14] for their imaging application. However, extensive tolerance studies of the RF system have not yet been performed which can be of great relevance for practical use of the system. To that end, we focus on the sensitivity of the RF system to deviation from the ideal magnification required to form aberration free volumetric images.

*Remote Focusing Principle*

For a lens, the defocus function describes the phase of the wavefront when a point source on the optical axis is shifted away from the focal plane. For a low NA lens, the phase term, $\psi$, can be written as a quadratic function:

$$\psi = nkz\,\rho^2\ \sin^2\alpha\,. \tag{1}$$

Here $n$ is the refractive index of the immersion medium of the lens, $k$ is the wavenumber equal to $\frac{2\pi}{\lambda}$ and $z$ is the axial shift from the focal plane of the lens. The normalised pupil radius, $\rho$, is defined as $\frac{\sin\theta}{\sin\alpha}$ where $\theta$ is the angle of the ray leaving the sample and $\alpha$ is the maximum acceptance angle of the lens. This defocus term can be easily compensated by shifting the detector till the image of the point source is in focus. However, for high-NA lenses, defocus is described by a spherical function[15]:

$$\psi = nkz\sqrt{1-\rho^2\ \sin^2\alpha}\,. \tag{2}$$

The term in the square root can be expanded to give higher orders of $\rho$ which is observed as depth dependant spherical aberration. Point sources outside of the focal plane (away from the objective) generate positive spherical aberration, with points inside focus generating negative spherical aberration. Any remote system used for refocusing high-NA objectives needs to produce equal and opposite amounts of the phase term described by eqn. 2 to compensate for the spherical aberration. An RF system does this exactly by introducing a matching high-NA lens in the optical path.

Fig.1 shows the optical layout of an RF system in the unfolded geometry. It consists of three infinity-corrected microscopes (S1, S2 and S3) in series. The first two tube lenses (L1 and L2) forming the relay optics (4f-system) are in telecentric alignment. S1 consists of the imaging objective, O1, which is closest to the sample being imaged and remains stationary. S2 is placed back to back with S1 so that it demagnifies the intermediate image to form an aberration free remote volume around the focal plane of the refocusing objective O2. A third microscope, S3, containing the reimaging objective O3 relays individual planes from the remote volume to the detector. This arrangement can also be configured in the folded geometry (schematic of experimental setup in folded geometry shown in Fig. 4) where O2 is reused as the reimaging objective by axially translating a mirror at its focal plane. As mirrors have lower inertia than objectives, this configuration allows for fast scan rates. However, in the folded geometry, half of the fluorescence signal is lost due to the presence of a polarising beamsplitter (placed immediately before O2). In this paper, subscripts 1, 2 and 3 denote parameters relating to the three microscopes shown in Fig. 1.

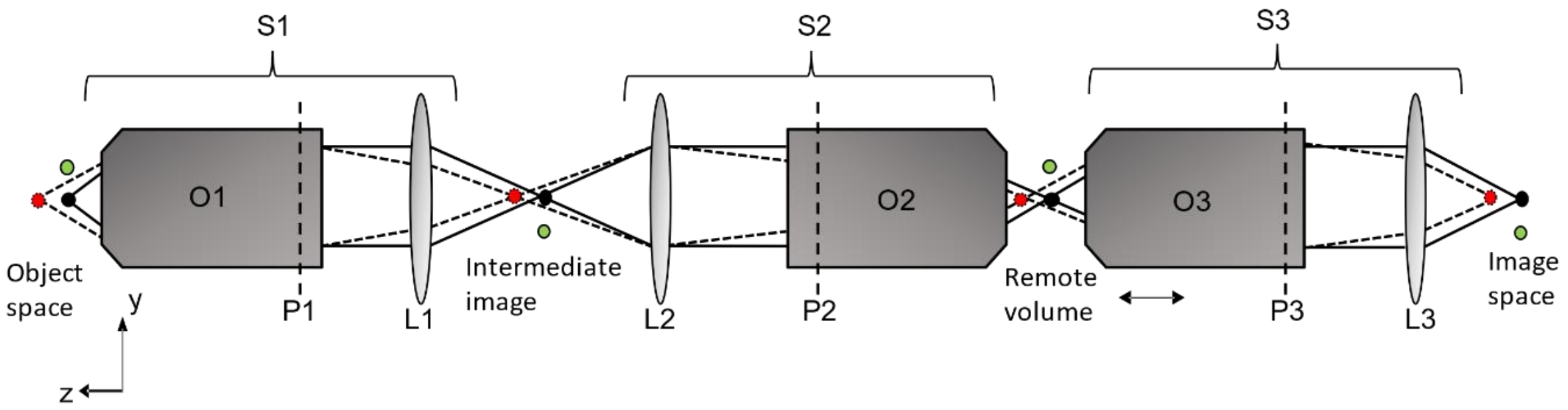

Fig. 1. An RF system in the “unfolded” geometry with three microscopes S1, S2 and S3 in series. An intermediate image is formed with magnification $M_{S1}$. The remote volume has a uniform magnification of $M_{S1}M_{S2} = \frac{n_1}{n_2}$. The final image formed by S3 on the detector has a magnification of $M_{S1}M_{S2}M_{S3}$. O3 is translated axially to image different depths of the remote volume. The vertical dashed lines on the objectives signify the position of the exit pupil plane (P) for each objective.

The formation of the aberration free remote volume can be understood by looking at two design conditions that are used to characterise lenses. The sine condition ensures that all points on a single plane perpendicular to the optical axis are imaged stigmatically (with no aberrations)[16]. Microscope objectives are designed using the sine condition which allows distortion-free imaging of laterally (x-y) shifted points on the focal plane. Complimentary to the sine condition, the Herschel condition allows for stigmatic imaging of points lying on the optical axis but displaced axially. As we require the formation of a volume that has no distortion laterally or axially, the RF system needs to simultaneously follow both the sine and Herschel condition. To do this the magnification of the system should be equal to the ratio of the refractive indices of the immersion media in the object and image space ($n_1$ and $n_2$ respectively)[17]. As objectives are designed to provide very high magnifications, the image formed by S1 is demagnified by S2 to form the remote volume having uniform magnification of

$$M_{RF}^{Id} = M_{S1}M_{S2} = \frac{n_1}{n_2}. \quad (3)$$

Where the magnification of the microscopes S1 and S2 are defined as

$$M_{S1} = \frac{M_1\, f_{L1}}{f_{L1,nom}}, \quad (4)$$

$$M_{S2} = \left[\frac{M_2\, f_{L2}}{f_{L2,nom}}\right]^{-1}. \quad (5)$$

$f_{L1,nom}$ and $f_{L2,nom}$ are the nominal focal lengths of the tube lenses and $M_1$ and $M_2$ are the nominal magnifications of O1 and O2 respectively. $f_{L1}$ and $f_{L2}$ are the focal lengths of the lenses used in the relay system.

For the remote volume to have a magnification defined by eqn. 3, it requires the magnification of the relay lenses to be

$$M_{4f}^{Id} = \frac{f_{L2}}{f_{L1}} = \frac{n_2 M_1}{n_1 M_2}. \quad (6)$$

For the simplest case of having identical objectives for O1 and O2 and the same immersion media for both, $M_{4f}^{Id}$ will be equal to 1. However, for biological applications, O1 is chosen such that the refractive index of the immersion media matches with that of the sample. O2 is preferred to be an air spaced objective so that inertial artefacts during refocusing can be avoided. This leads to reduced flexibility in the choice of lenses for the relay system. To get the maximum axial extent of aberration free imaging (dynamic range) requires the relay lenses to closely follow eqn. 6 which will lead to the RF system having the ideal magnification defined by eqn. 3. Deviation leads to breaking the Herschel condition which again results in the introduction of spherical aberration terms and reduction in the dynamic range.

In the following sections we determine the sensitivity of the dynamic range to the choice of lenses, L1 and L2. We first build a computational model that can predict the amount of spherical aberration present at each depth for different amounts of magnification mismatch. We then validate this model against experimental measurements of pupil plane aberrations in a folded remote focusing system using a Shack-Hartmann wavefront sensor. Finally, we use the simulation to quantify the sensitivity of the dynamic range to the magnification mismatch.

## Methods

*Remote Focusing Model*

As magnification mismatch introduces aberrations into the RF system at defocussed positions, this can be represented as phase variations in the wavefront at the pupil plane. Following the RF theoretical model built by Botcherby, we consider a point source on the optical axis shifted by a distance '$z$' from the focal plane, which in turn is located a distance '$f$' from the lens. In the condition that $z \ll f$, the generalised phase at the pupil planes P1 and P2 are given by[7]

$$\psi_1 = n_1 k \left\{ f_1 \left( 1 - \frac{2}{f_1} (1 - \rho_1^2 sin^2 \alpha_1)^{\frac{1}{2}} + \frac{{z_1}^2}{f_1^2} \right)^{\frac{1}{2}} - f_1 \right\}, \tag{7}$$

$$\psi_2 = n_2 k \left\{ f_2 \left( 1 + \frac{2}{f_2} (1 - \rho_2^2 sin^2 \alpha_2)^{\frac{1}{2}} + \frac{{z_2}^2}{f_2^2} \right)^{\frac{1}{2}} - f_2 \right\}. \tag{8}$$

Here $f_{1,2}$ is the front focal length of the objectives (O1 and O2) and can be calculated by dividing the focal length of the tube lens by the magnification of the microscope. The parameters $z_1$ and $z_2$ are the distances of the point sources away from the focal planes of O1 and O2 respectively. The normalised pupil radius, $\rho$, ranges from 0 to 1 from the centre to the edge of the pupil (Fig. 2a).

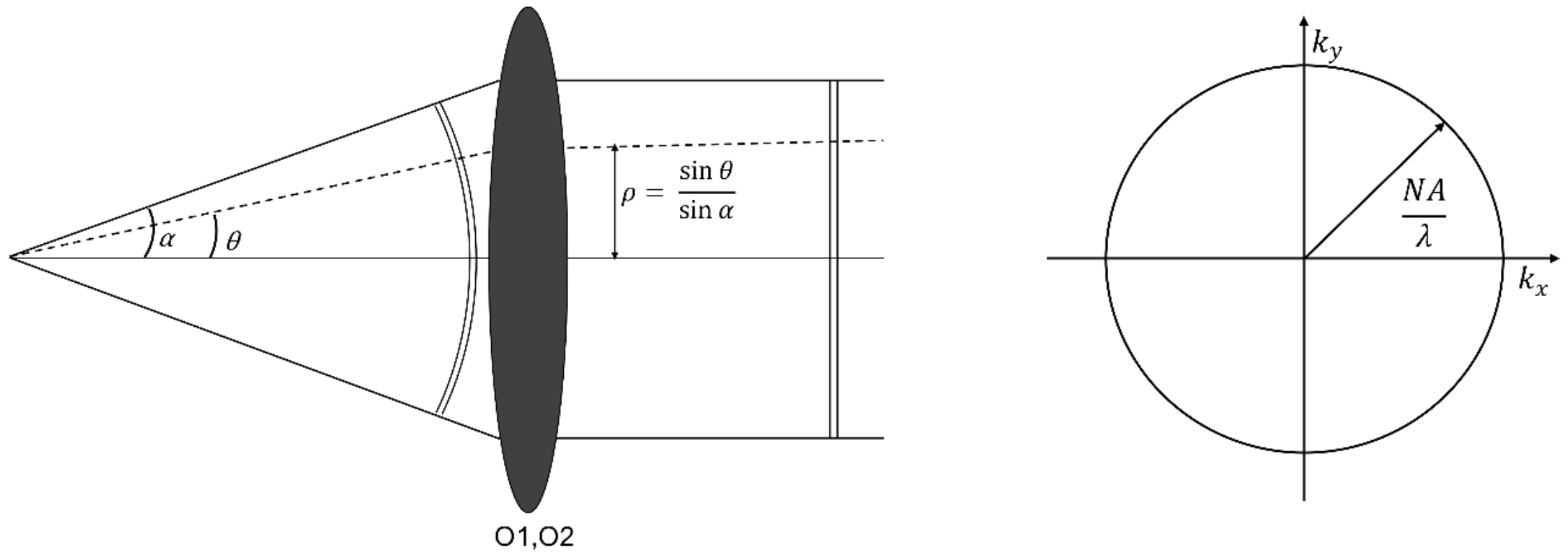

(a) (b)

Fig. 2. (a) Schematic describing the relation between ray angles and normalised pupil radius (b) Schematic showing the radial extent of the pupil in frequency space.

The phase of the wavefront forming the remote volume, $\Delta\psi_{RF}$, is calculated by taking the sum of the phase terms defined by eqns. 7 and 8 which we approximate and rewrite as:

$$\psi_1 = n_1 k \left(1 - z_1(1-\rho_1^2 sin^2\alpha_1)^{\frac{1}{2}} + \frac{z_1^2(1-\rho_1^2 sin^2\alpha_1)}{2f_1}\right), \tag{9}$$

$$\psi_2 = n_2 k \left(1 + z_2(1-\rho_2^2 sin^2\alpha_2)^{\frac{1}{2}} + \frac{z_2^2(1-\rho_2^2 sin^2\alpha_2)}{2f_2}\right), \tag{10}$$

$$\Delta\psi_{RF} = \psi_1 + \psi_2 \,. \tag{11}$$

We introduce a factor, β, which is the ratio of the actual relay lens magnification to the ideal magnification.

$$\beta = \frac{M_{4f}}{M_{4f}^{Id}} \tag{12}$$

When β=1, the mapping is ideal and both Herschel and sine conditions are satisfied (Eqn. 3). The function of the relay lenses is to ensure that the wavefront mapped onto the pupil plane of O2 is equal and opposite to that formed in the pupil plane of O1. This results in the spatial frequencies being accurately mapped leading to $\rho_1 \sin\alpha_1 = \rho_2 \sin\alpha_2$, for all rays, cancelling the linear $z$ terms in eqns. 9 and 10. Therefore, the wavefront formed by a point object at axial displacement $z_1$ is stigmatically imaged by O2 at $-\frac{n_1}{n_2}z_2$. However, as $z_1$ increases, the $z_i^2$ terms relating to higher order spherical aberration add up to contribute to $\Delta\psi_{RF}$, limiting the theoretical dynamic range of an ideal RF system. In a non-ideal system, where $M_{4f}$ is not equal to $M_{4f}^{Id}$, non-cancellation of the linear $z$ terms results in increased amounts of spherical aberration even for small shifts in $z_1$. β>1 and β<1 signifies over-magnification and under-magnification by the relay lenses respectively.

*Computational Model*

We characterise the sensitivity of the RF system to magnification mismatch by quantifying the amount of spherical aberration generated by a point source translated by a distance $z$ from the focal plane of O1. This can be done by using equations 9 and 10 to calculate the phase at the pupil plane of O1 and O2 for different $\beta$ and deriving the resulting $\Delta\psi_{RF}$. The pupil plane is described using spatial frequency coordinates $k_x$ and $k_y$ (Fig. 2b). This pupil plane is subdivided into 2N * 2N regions, such that the smallest increment in $k_x$ or $k_y$ is defined by:

$$\gamma_x = \gamma_y = \frac{NA}{\lambda N}. \tag{13}$$

The $\boldsymbol{k}$ vector within the pupil plane is therefore:

$$\boldsymbol{k} = \begin{pmatrix} k_x \\ k_y \end{pmatrix} = \begin{pmatrix} m\gamma_x \\ n\gamma_y \end{pmatrix}. \tag{14}$$

For each location in the pupil plane $(m, n)$ , we calculate the $\sin\theta_1$ value of the corresponding ray to be

$$\sin\theta_1(m,n) = \frac{\lambda\,\gamma\,\sqrt{k_x^2 + k_y^2}}{n_1} \ . \tag{15}$$

From the $\sin\theta_1(m,n)$ values we can then calculate $\rho_1(m,n)$ and $\cos\theta_1\,(m,n)$ as:

$$\rho_1(m,n) = \frac{\sin\theta_1(m,n)}{\sin\alpha_1}, \tag{16}$$

$$\cos\theta_1\,(m,n) = \sqrt{(1 - \rho_1^2 sin^2\alpha_1)}\,. \tag{17}$$

where $\sin\alpha_1 = \frac{NA_1}{n_1}$. To map between the two pupil planes, we use the relation:

$$\sin\theta_2(m,n) = \sin\theta_1(m,n) * \beta\,. \tag{18}$$

If $\beta$ =1, $\sin\theta_2(m,n) = \sin\theta_1(m,n)$ . For $\beta \neq 1$, the mismatch in frequency space is reflected in the final phase of the wavefront, $\Delta\psi_{RF}$, introducing aberrations in the remote volume.

To impose the finite extent of allowed spatial frequencies we define a circular mask in the pupil plane (Fig. 2b) as

$$\text{Pupil Mask} = \begin{cases} 1 & \sqrt{k_x^2 + k_y^2} \leq \dfrac{n_1 \sin\theta_{max}}{\lambda} \\ 0 & \text{otherwise.} \end{cases} \tag{19}$$

In eqn. 19, $\sin\theta_{max}$ is the limiting aperture of the RF system. For the spatial resolution of an RF system to be defined by the NA of O1, $\sin\alpha_2$ should be greater than or equal to $\sin\alpha_1$. This ensures that O2 does not act as an aperture stop in the RF system. This important RF design condition is considered to be true in the simulation and the pupil mask for a $\beta = 1$ system is defined as $\sin\theta_{max} = \sin\alpha_1$. For non-ideal conditions, the pupil mask is calculated for the objective limiting the ray angles by looking at both the forwards and backwards geometry of the RF system.

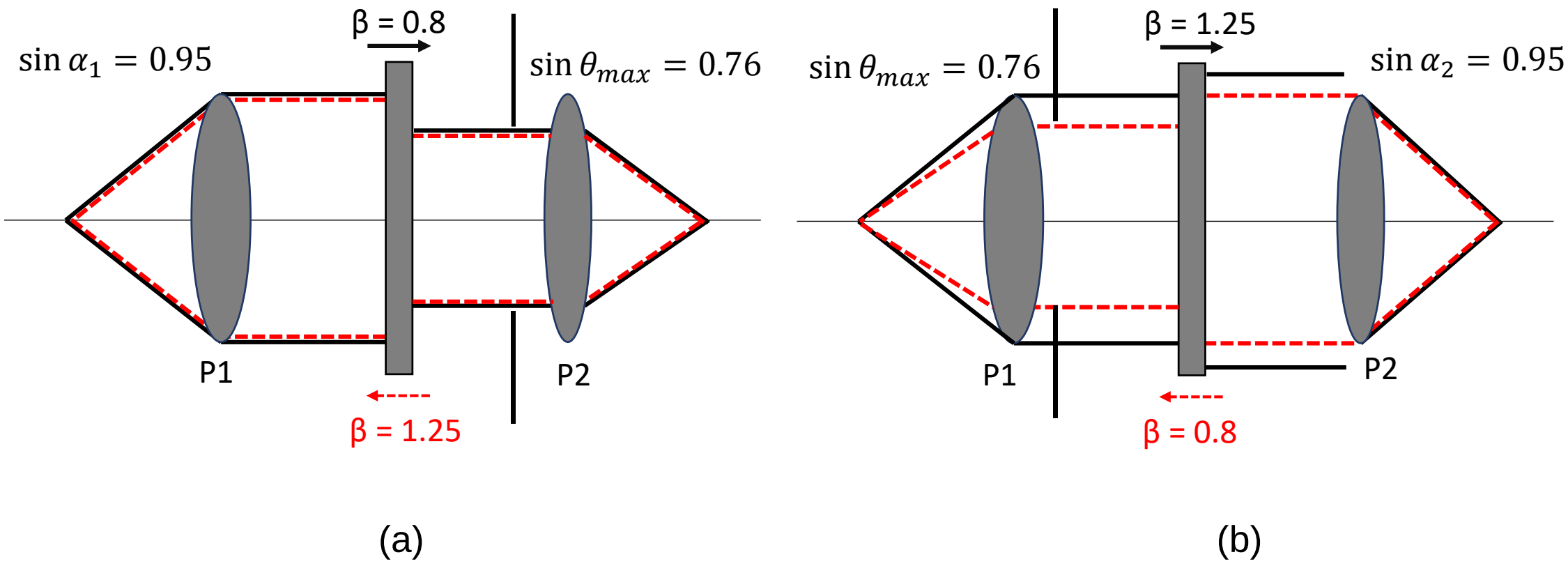

Fig. 3. Shows the selection of the limiting aperture for two identical objectives ($\sin\alpha_1 = \sin\alpha_2 = 0.95$) with non-ideal relay lens magnifications (a) $\beta = 0.8$ and (b) $\beta = 1.25$. P1 and P2 indicate the pupil planes of the objectives. The solid black and red dashed lines show the forward and backward ray traces respectively.

For β<1, O2 acts as an aperture stop to give $\sin\theta_{max} = \sin\alpha_1 * \beta$. For β>1, O1 acts as the aperture stop to give $\sin\theta_{max} = \frac{\sin\alpha_1}{\beta}$. The pupil mask (eqn. 19) multiplied by the total phase term (eqn. 11) gives the final form of the wavefront forming the remote volume in the RF system. The simulation was performed using MATLAB software and the code is made available here: https://github.com/sharika-mohanan/RF_System.git

*Zernike Terms*

In order to obtain the contribution of spherical aberration terms at defocused positions, $z$, we decompose the pupil phase into radially symmetric set of Zernike polynomials[18]. As spherical aberration also introduces defocus into the imaging system, it shifts the refocused image by $\delta z$. This displacement aberration can be optically compensated and is therefore subtracted from $\Delta\psi_{RF}$. The defocus function $\psi_d$ and the defocus coefficient $\delta z$ are taken in the same form as eqns. 20 and 21 in reference 7 to give the final form of the wavefront

$$\psi'_{RF} = \Delta\psi_{RF} - \delta z \psi_d. \tag{20}$$

$\psi'_{RF}$ can then be expanded as a series of Zernike polynomials $Z_p^q$:

$$\psi'_{RF} = nk \sum_{p=0}^{\infty} \sum_{q=-p}^{n} C_p^q Z_p^q \,. \tag{21}$$

Here $p$ is the axial order and $q$ is the azimuthal order of the expansion terms. The polynomials, $Z_p^q$, are orthogonal to each other over a unit circle and $C_p^q$ are the expansion coefficients which quantify the contribution of each aberration mode to the total phase. As the simulation models a point source shifted along the optical axis, the azimuthal terms can be ignored. This leaves the terms relating to $Z_p^0$ which are the rotationally symmetric aberrations. $\psi'_{RF}$ was expanded to the first 25 terms and the fitting was performed using the MATLAB zernike_coeffs function[19]. In the expansion basis set, the polynomial associated with first order spherical aberration is

$$Z_4^0(\rho) = \sqrt{5}(6\rho^4 - 6\rho^2 + 1)\,. \tag{22}$$

The corresponding first order spherical aberration coefficient, $C_4^0$, was extracted for a range of $z$ and compared with the experimentally derived values.

*Strehl Ratio*

We characterised the tolerance of the RF system to magnification mismatch by measuring the change in the dynamic range for different $\beta$. To do this, we measured the Strehl ratio across $z$ for different $\beta$ values. For an RF system, the Strehl ratio ($S$) is defined as the ratio of the maximum intensity of the image of the point source at $z$ to that at $z = 0$ (focal plane). It can also be written as[20]

$$S = e^{-\langle (\psi'_{RF} - \overline{\psi'_{RF}})^2 \rangle}\,. \tag{23}$$

An unaberrated wavefront has a Strehl ratio of 1. Due to its dependence on the variance of the wavefront, increased amounts of aberrations reduces the Strehl ratio. Similar to the previous section, we use $\psi'_{RF}$ to calculate $S$ as the presence of defocus terms increases the variance which would underestimate the maximum attainable dynamic range. As explained in reference 18, Strehl ratio of 0.8 and above is considered nominal for perfect imaging and therefore sets the bounds of the dynamic range for a given $\beta$.

*Experimental Verification*

The computational model was verified experimentally by constructing an RF system in folded geometry (Fig. 4). A pair of 0.95 NA 40x dry objectives (UPLSAPO40X2, Olympus) were used as the imaging and reference objectives O1 and O2 respectively. A collimated laser beam (532nm, CPS532, Thorlabs) was expanded to 10 mm diameter to overfill the back aperture of O1. The focal spot formed by O1 approximated as a point source for the RF system. Mirror R1, mounted on a linear stage (PT1A/M, Thorlabs) was translated axially across the focal plane. When R1 is translated by a distance $z_1$, the optical path length changes by a factor of 2, changing the object position by $z = 2z_1$. The corresponding refocused image was formed by O2 at $z = -2z_2$ as $n_1 = n_2$ .

The relay lenses L1 and L2 map the pupil planes of O1 and O2 together and were placed in telecentric alignment. To ensure that the aberrations arising due to the misalignment of the mirrors were kept minimum, the mirrors were translated across their axial range and the reflected beam was checked to be centred across the optical layout using a pinhole.

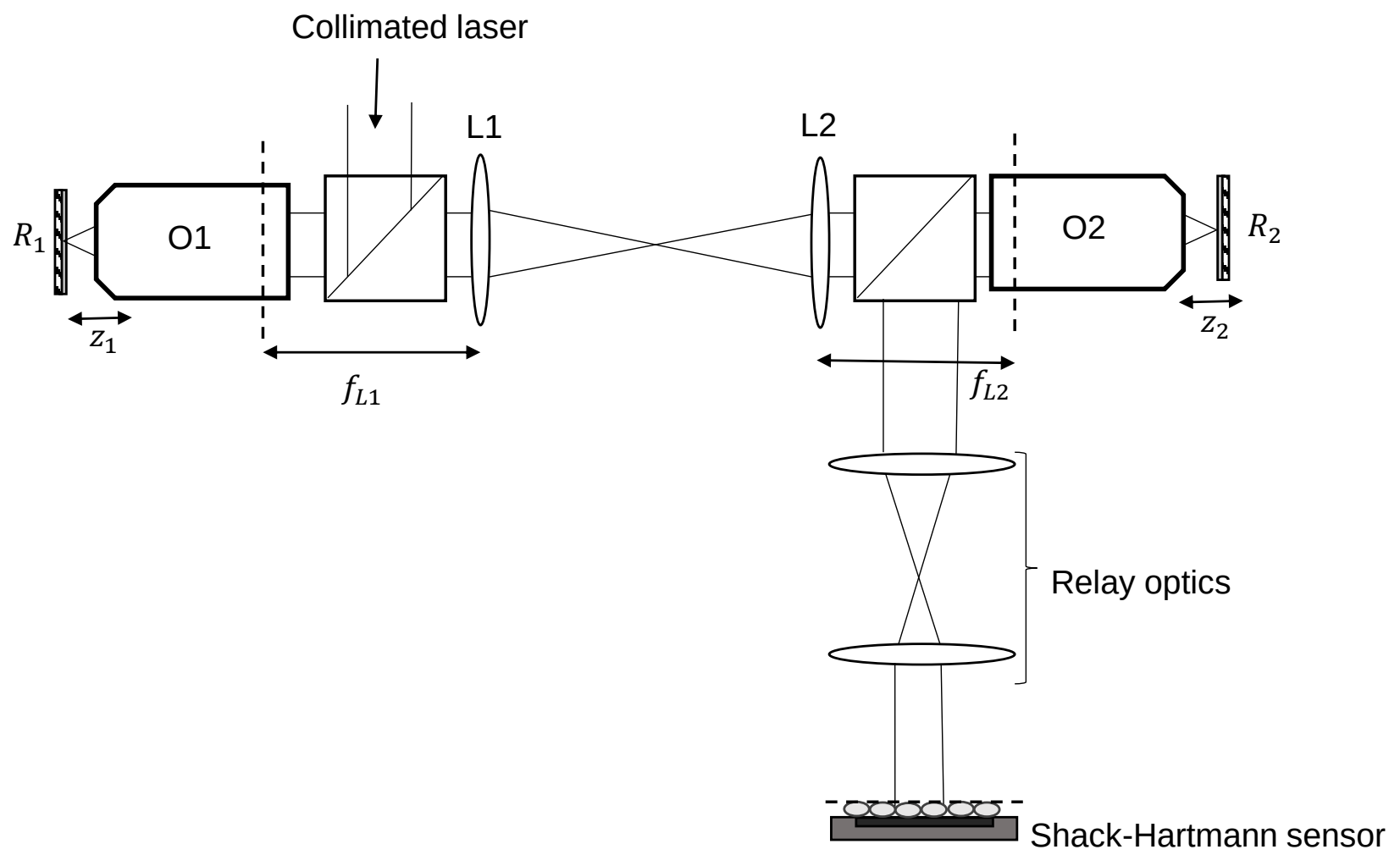

Fig. 4. RF system aligned in folded geometry used for computational model verification.

Three achromat lens pairs were chosen for L1 and L2 to provide the following focal length combinations $(f_{L1}, f_{L2})$, (in mm) : (125, 125), (100, 125) and (125, 100). This gave $\beta$ values of 1,0.8 and 1.25 to reflect ideal, under-magnified and over-magnified configurations. To measure the amount of aberrations, present in the final wavefront forming the remote volume, the pupil plane of O2 was made conjugate to the lenslet array of a Shack-Hartmann wavefront sensor using another pair of relay lenses.

*Shack-Hartmann sensor*

The Shack-Hartmann (SH) wavefront sensor was built using a microlens array (MLA300-14AR-M, Thorlabs) and a CMOS camera (UI-3240LE-M-GL, IDS). The camera sensor was placed at the focal plane of the lenslet array. The sensor was divided into grids whose dimensions were defined by the lenslet diameter (300 μm). For a wavefront with no aberrations, each lenslet focuses the light to the centre of the grid. For an aberrated wavefront, the displacement of the focal spot within each grid was measured by finding the spot centre using centroiding algorithms. This spot shift is directly related to the average local slope of the wavefront at each lenslet. The maximum Optical Path Difference (OPD) that can be measured by this SH sensor was calculated to be 3 μm. Modal reconstruction using Zernike polynomials as the basis set[21] was used to reconstruct the final wavefront (Fig. 5). The SH sensor was then calibrated using a pure spherical wavefront emitted by a single mode fibre[22].

When aligning the RF system (Fig. 4), the SH sensor was used to ensure that the incoming collimated beam had minimal aberrations. The same was checked for the RF system without O1 and O2 present. Once the objectives were in place, the wavefront taken at $z = 0\ \mu\text{m}$ was used as the reference wavefront to minimise contributions from any misalignment. For each displacement of R1, R2 was translated until the defocus term was completely cancelled ($C_2^0 = 0$). As the contribution of the second order spherical aberrations was insignificant, it was not used for further analysis. The coefficient of the first order spherical aberration was extracted from the final reconstructed wavefront to compare with the computational model results.

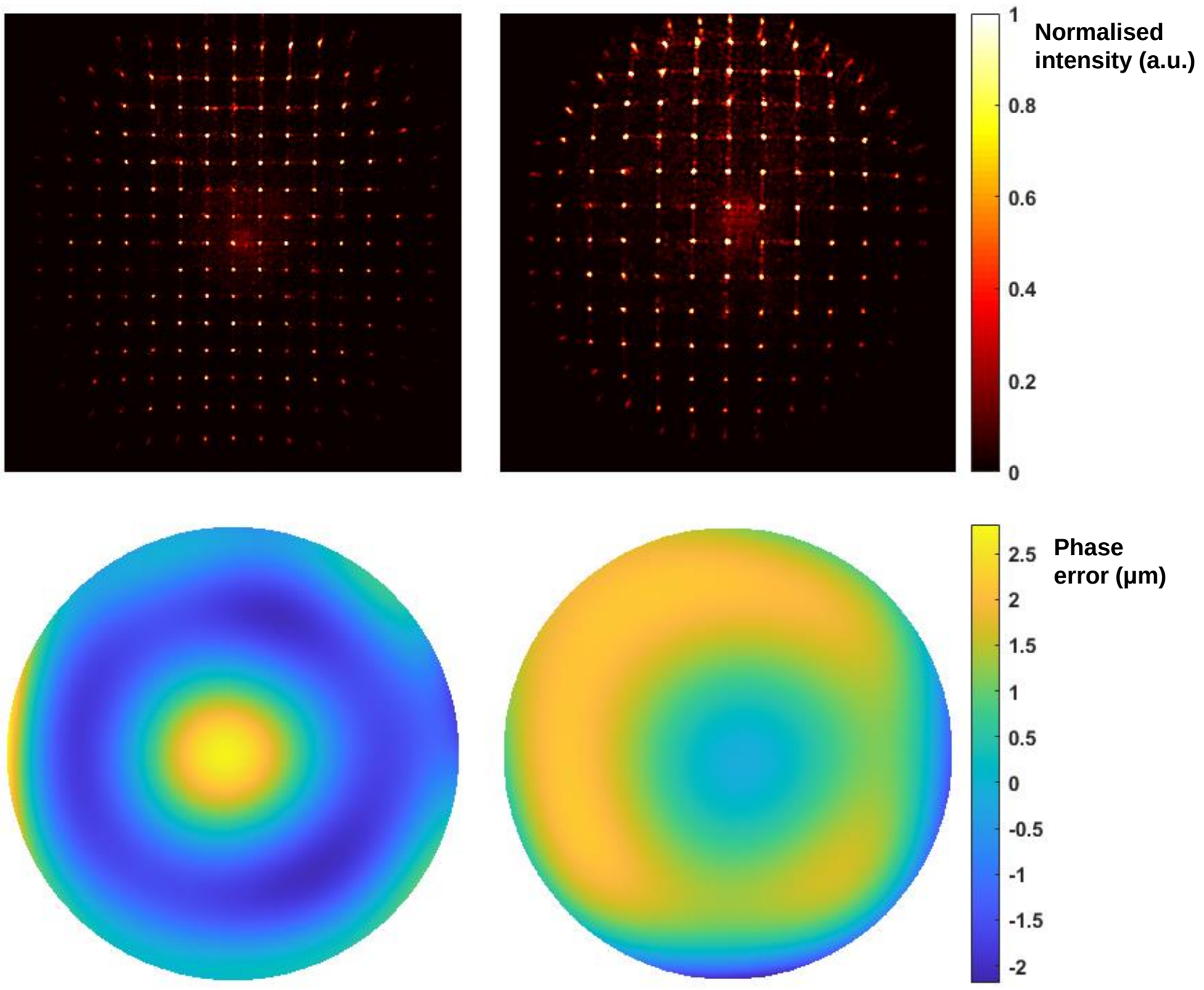

Fig. 5. (Top) Shack-Hartmann images taken for $\beta = 1.25$ (left) and $\beta = 0.8$ (right) at $z$ = 50 μm. (Bottom) The corresponding reconstructed wavefronts.

*Axial Point Spread Function Measurement*

The computational model was also applied to a second RF system. This system was built in unfolded geometry using a 1.15 NA 40x water immersion objective as the imaging objective O1. A pair of 0.95 NA 40x dry objectives (UPLSAPO40X2, Olympus) were used for O2 and O3 (see reference 10 for details). L1 and L2 was changed from 180 mm – 135 mm to 180 mm – 140 mm to change $\beta$ from 1 to 1.04. 100 μm fluorescent beads (F8803, ThermoFisher, Excitation:505 nm, Emmision:515 nm) suspended in 2% solution of agarose was used as the sample. These sub-resolution beads act as point sources and were sparse enough within the sample to allow imaging without sectioning. O3 was translated axially every 0.2 μm using a piezo stage (Q-545.140, Physik Instrumente) across a 400 μm range. The image stacks were captured on an sCMOS camera (Zyla 4.2, Andor Technology, Oxford Instruments). The beads at different depths were analysed using PSFj software[23] and the fitted XZ Point Spread Function (PSF) profiles were used for further qualitative analysis.

## Results and Discussion

*Effect of Magnification Mismatch*

The experimental parameters of the folded RF system (Fig. 4) were fed into the computational model. To ensure that the sampling of the phase at the pupil plane did not introduce aliasing effects, we measured the first order spherical aberration coefficient $C_4^0$ (Fig. 6.) as a function of the image size N. $C_4^0$ was calculated for $z = 80\ \mu m$, β = 1.25 where high amounts of spherical aberration was observed. The curve in Fig. 6 flattens out for N > 500 pixels and all the simulations were performed keeping N = 1280 pixels.

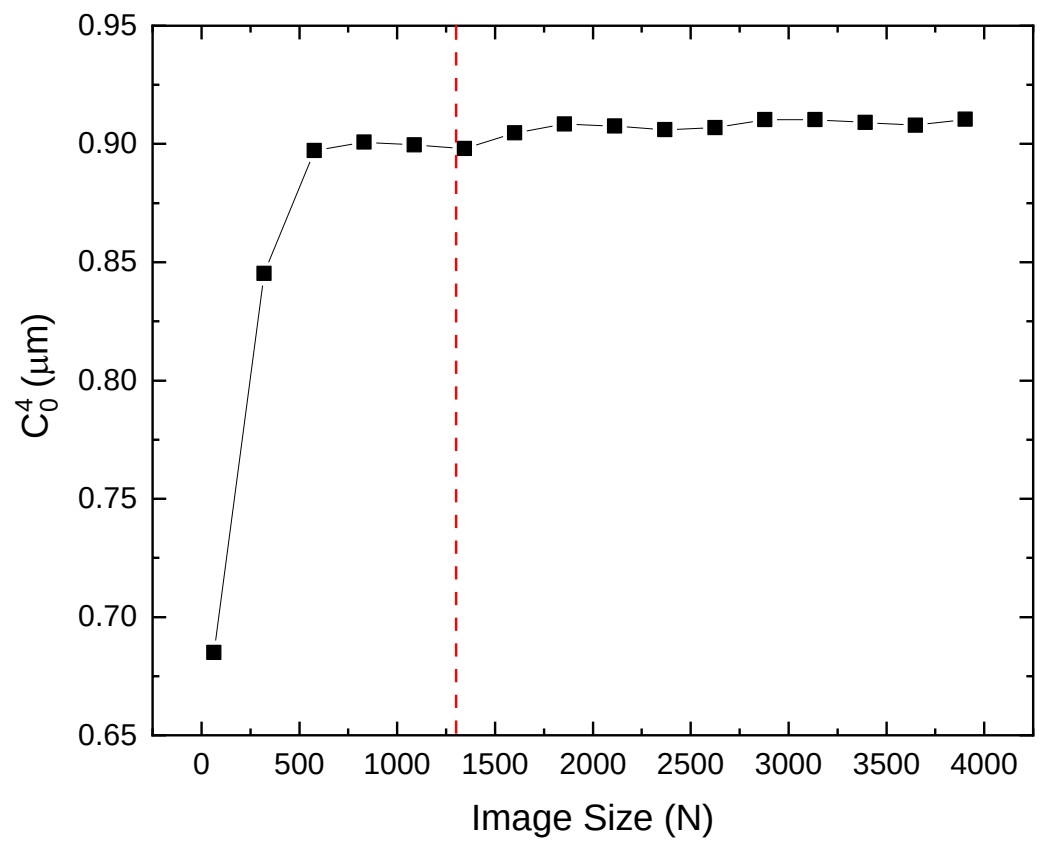

Fig. 6. First order spherical aberration coefficient ($C_4^0$) calculated for different image size (N). The vertical red dashed line shows the image size used in the computational modelling of the pupil plane.

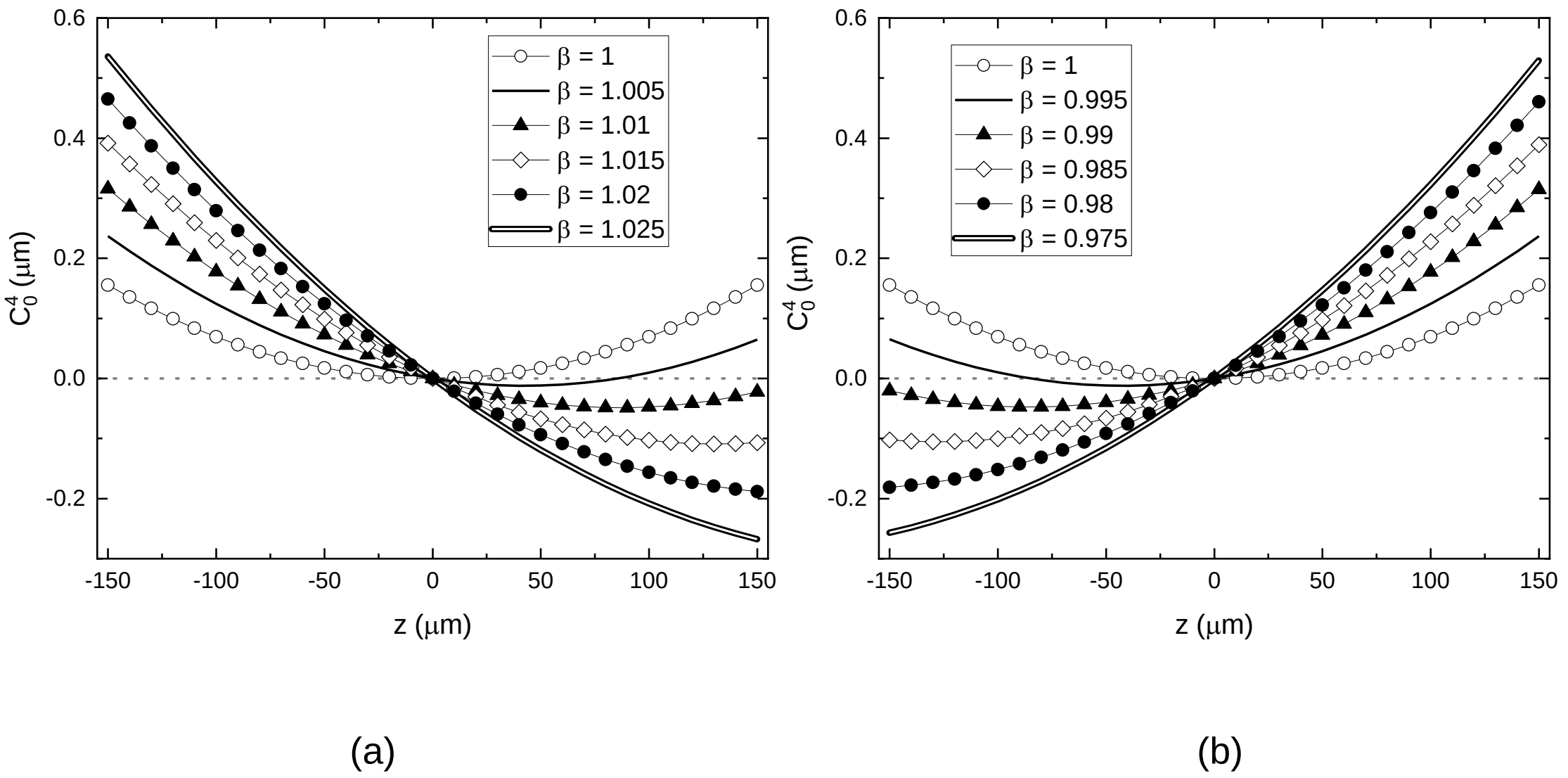

Fig. 7. Comparison of line profiles across $z$ for small changes in magnification for (a) over-magnified ($\beta > 1$) and (b) under-magnified ($\beta < 1$) conditions.

To assess the increase in spherical aberration with depth we look at the change in first order spherical aberration coefficient ($C_4^0$) with $z$. The simulation was performed using two air lenses for O1 and O2 (0.95NA 40x, same as the experimental system in folded geometry) while changing $\beta$ to reflect over-magnified and under-magnified conditions. For small variations in $\beta$ (less than a 5% change), there is an equivalent contribution from both the quadratic $z^2$ term and the linear $z$ term from eqns. 9 and 10 resulting in the skewed line profiles. Another observation is that for small changes in $\beta$, $C_4^0$ crosses zero at two $z$ positions. Once at the nominal focal plane of O2 ($z = 0$) and another at a position shifted from this focal plane. This crossing is seen for $\beta = 1.005, 0.995$ in Fig. 7 at ~ 90 µm away from the focus. This should result in this non-ideal system seeing best imaging resolution at two positions in z. As $\beta$ increases, the linear $z$ terms start to dominate showing rapid increase in spherical aberration on either side of the focus.

The spherical aberration coefficient obtained from the computational model was compared with the corresponding term from the experimental system. Fig. 8 shows the change in $C_4^0$ with distance $z$ for the three different magnifications of the relay lenses. For ideal magnification ($\beta = 1$) we see reduced amounts of spherical aberration across $z$. For β ≠ 1 conditions, we observe the rapid increase in spherical aberration with distance from the focal plane. The deviation of the experimental values from the simulated results for $\beta = 1.25$ can be attributed to the presence of residual aberrations in the optical system due to alignment errors. The range over which the coefficient could be measured directly was limited by the dynamic range of the SH sensor. The measured coefficient can be seen to saturate close to the upper and lower limits of the axial range for both over and under magnified cases.

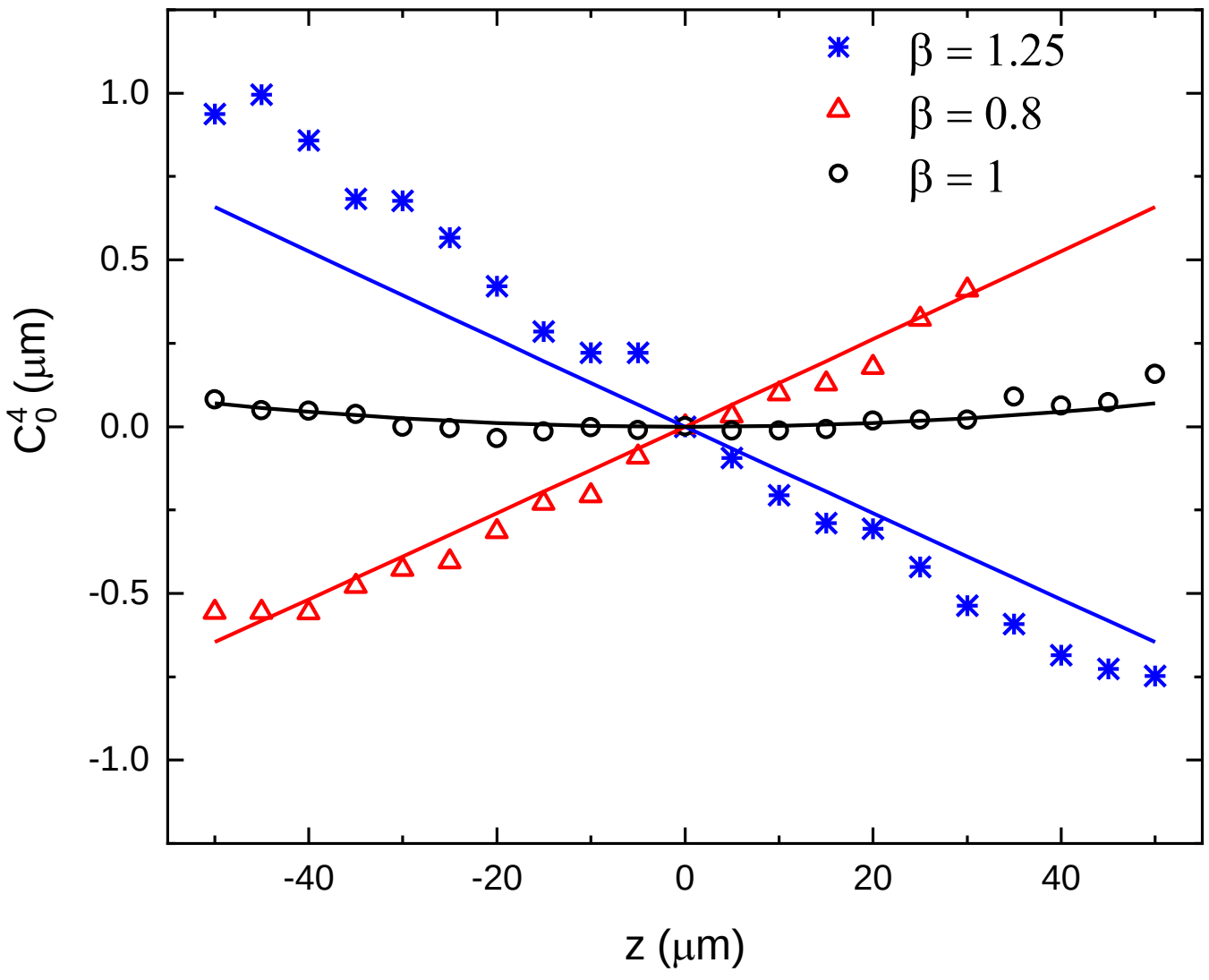

Fig. 8. Verification of first order spherical aberration terms obtained from experiment for ideal ($\beta = 1$), over-magnified ($\beta = 1.25$) and under-magnified ($\beta = 0.8$) RF systems. These are compared against their corresponding simulation results (solid lines).

*Decrease in Dynamic Range*

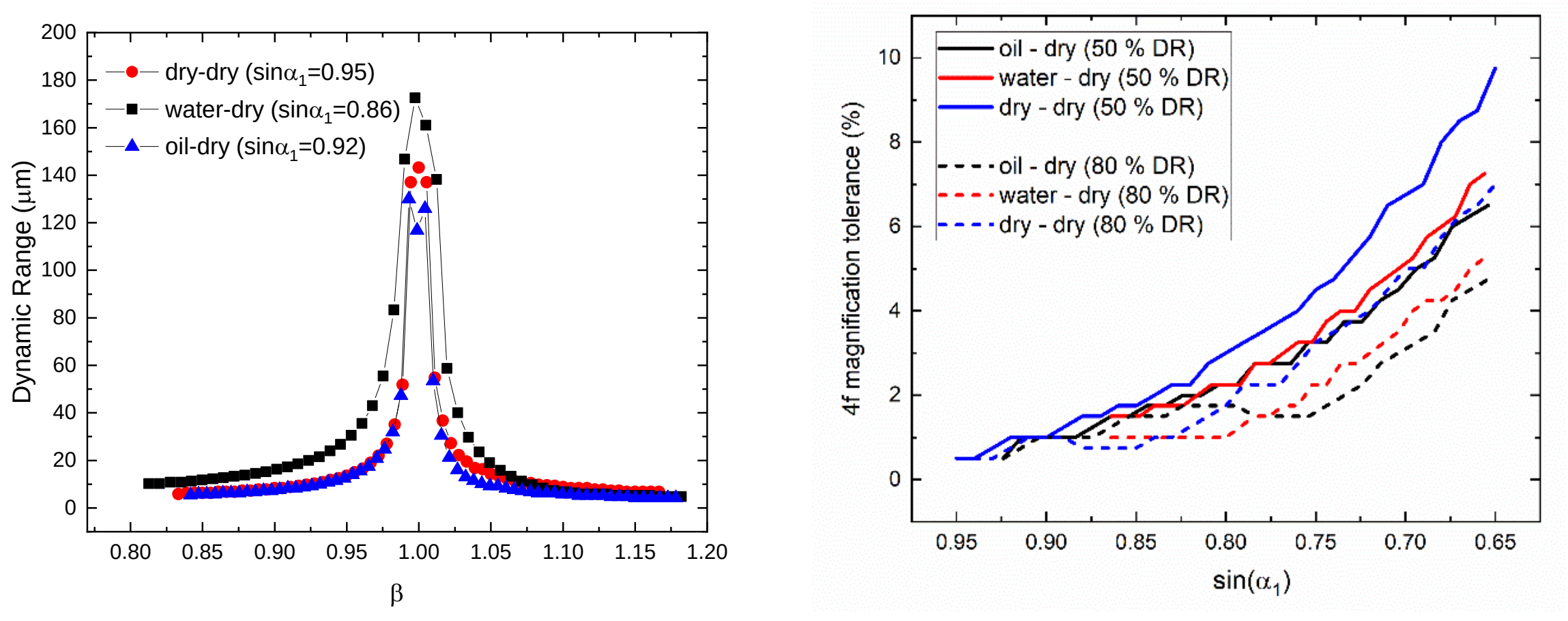

(a) (b)

Fig. 9. (a) Plot showing the decrease in dynamic range for three different O1-O2 objective pairs as a function of $\beta$. (b) The plot shows the magnification mismatch tolerance of the 4f system when trying to achieve 50% or 80% of the theoretical dynamic range (DR). The tolerance has been calculated for decreasing maximum acceptance angles of the imaging objective.

The Strehl ratio was calculated for $z$ ranging from -100 µm to +100 µm and the dynamic range was defined for the axial region having Strehl ratio greater than 0.8. Three combinations of O1 and O2 objectives were considered: 0.95 NA 40x dry – 0.95 NA 40x dry, 1.15 40x water immersion – 0.95 NA 40x dry and 1.4 60x oil immersion – 0.95 NA 40x dry. For all configurations, an approximate 1% change in $\beta$ shows a decrease in the dynamic range to at least half of the maximum value (Fig. 9a). As higher NA objectives generate spherical aberration at a much higher rate outside of the focal plane, it makes the corresponding RF system highly sensitive to small deviations from the ideal configurations. As seen in Fig. 9b, the tolerance of the RF system to magnification mismatch increases with decreasing $\sin\alpha_1$ values. Therefore, to reduce sensitivity of the system to magnification mismatch, one can stop down O1, which increases the dynamic range (Fig. 10) at the loss of the system resolution.

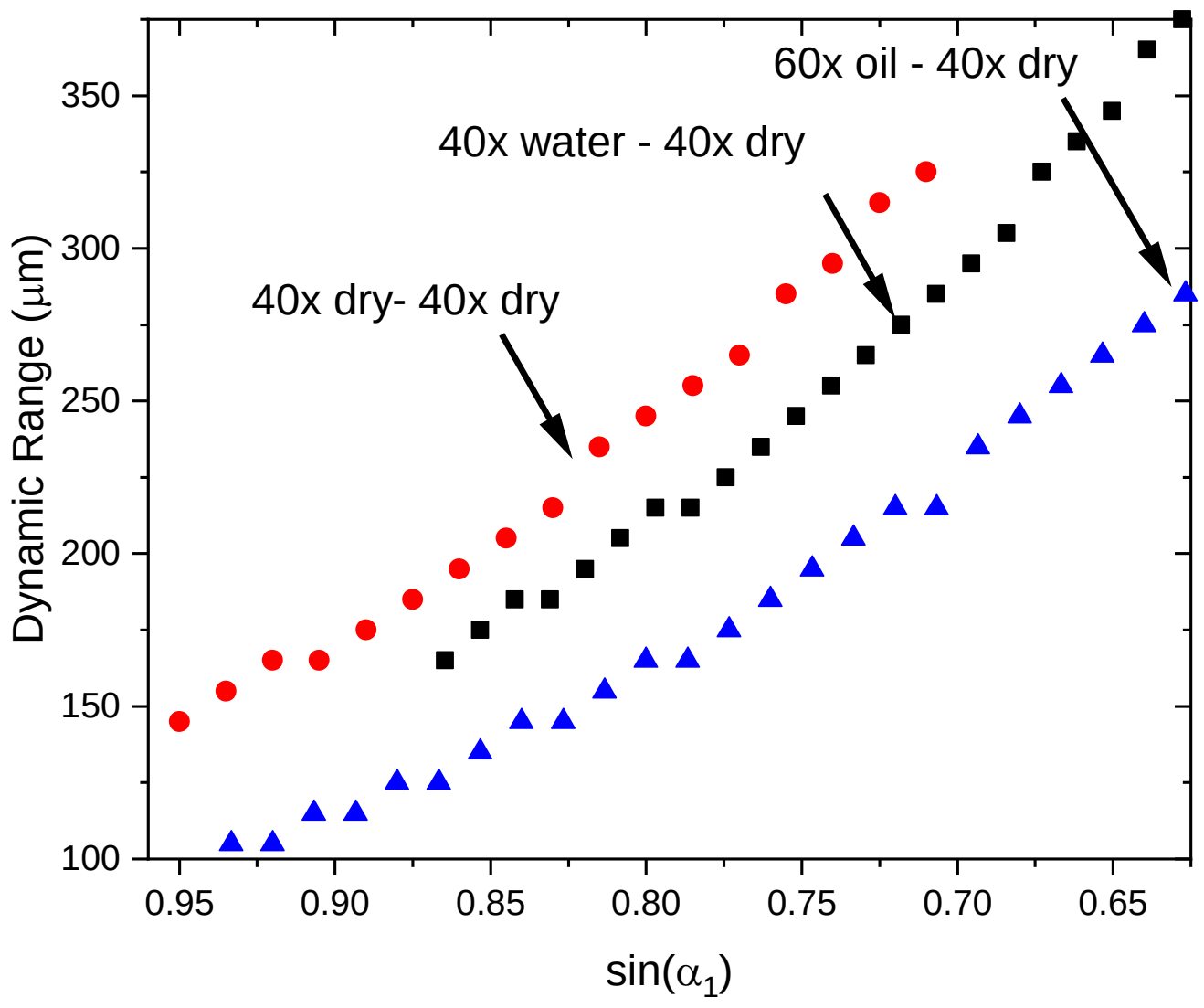

Fig. 10. The increase in dynamic range with decrease in the maximum acceptance angle of O1 is plotted for the same objective pairs as in Fig. 9.

*Empirical observations near ideal magnification*

It is also observed that for an ideal RF system there is complete cancellation of the linear $z$-dependant terms as is predicted by RF theory (eqns. 9-11). This is reflected in the symmetric profile of the curve corresponding to $\beta = 1$ in Fig. 8. This information can be used for characterising an RF system to check if the final remote volume has been formed with uniform magnification of $\frac{n_1}{n_2}$.

Using the second RF system in unfolded geometry, the axial PSF was measured using fluorescent beads to check the direction of elongation of the PSF 'tail' (Fig. 11). For an ideal system ($\beta = 1$) positive spherical aberration is observed on either side of the focal plane where the axial profile is elongated towards the refocusing objective O2. Whereas, in an over-magnified system ($\beta = 1.04$) the sign of the spherical aberration changes from positive to negative from $-z$ to $+z$.

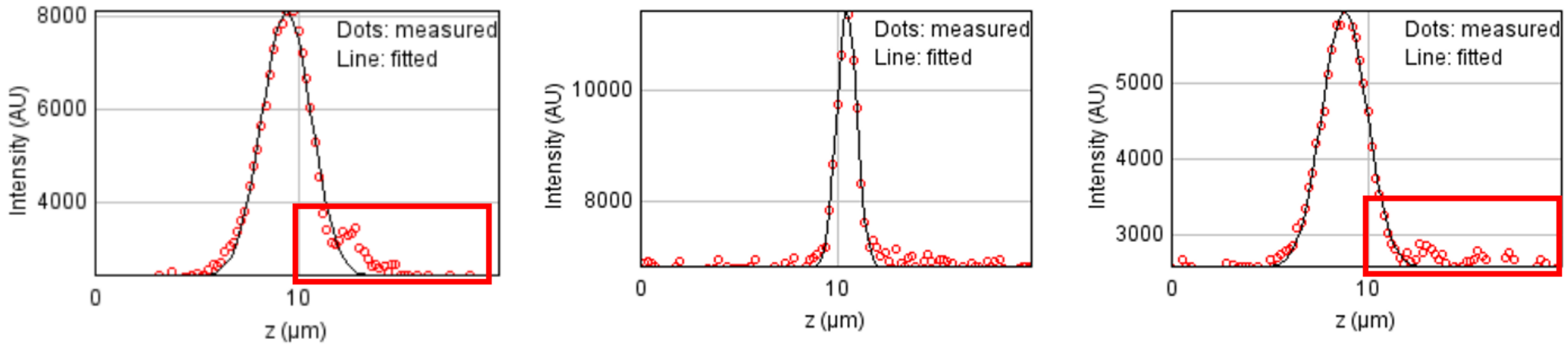

(a) (b) (c)

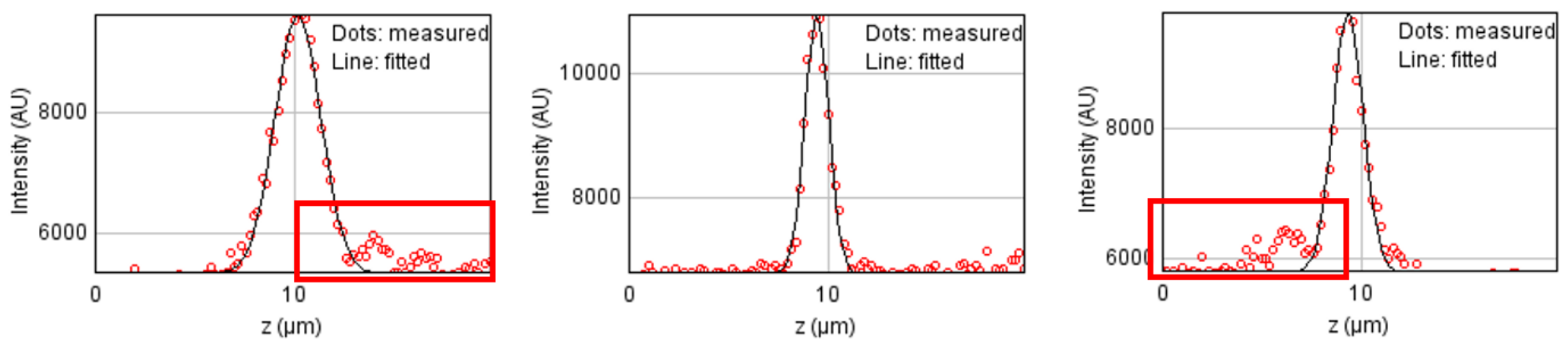

(d) (e) (f)

Fig. 11. Qualitative assessment of magnification in an RF system. Positive $z$ on the axis is towards the refocusing objective O2. Top row shows the line profile of the axial PSF taken for $\beta = 1$ at $z = -180\ \mu m$ (a), $z = 0\ \mu m$ (b) and $z = +180\ \mu m$ (c). Bottom row for $\beta = 1.04$ is imaged at $z = -140\ \mu m$ (a), $z = 0\ \mu m$ (b) and $z = +120\ \mu m$ (c).

## Conclusion

In this paper, we have presented a computational method to quantify the imaging properties of a remote focusing system. This model includes non-ideal RF configurations where the magnification of the system deviates from the ideal value. A folded RF system was built with three different relay lens magnifications to verify the computational model. The first order spherical aberration term obtained from the experiments was found to be in close agreement with the simulated results. The model was then extended to calculate the decrease in dynamic range for increasing magnification mismatch. It is predicted that a 1% change in magnification decreases the dynamic range to half of the maximum value. We also use the sign of the spherical aberration on either side of the focus for emperical verification of the remote volume reaching ideal magnification conditions.

## Acknowledgements

The authors would like to thank Prof. Christian Soeller for valuable feedback regarding the manuscript and Adam Westmacott (Olympus Keymed) for the generous loan of two UPLSAPO40X2 objectives.